\documentclass[11pt]{article}
\usepackage{a4,graphicx}

\usepackage[scaled]{helvet} 
\usepackage{courier} 
\usepackage{eulervm} 
\normalfont
\usepackage[T1]{fontenc}

\begin{document} 
\begin{figure}
\includegraphics[width=0.2\textwidth]{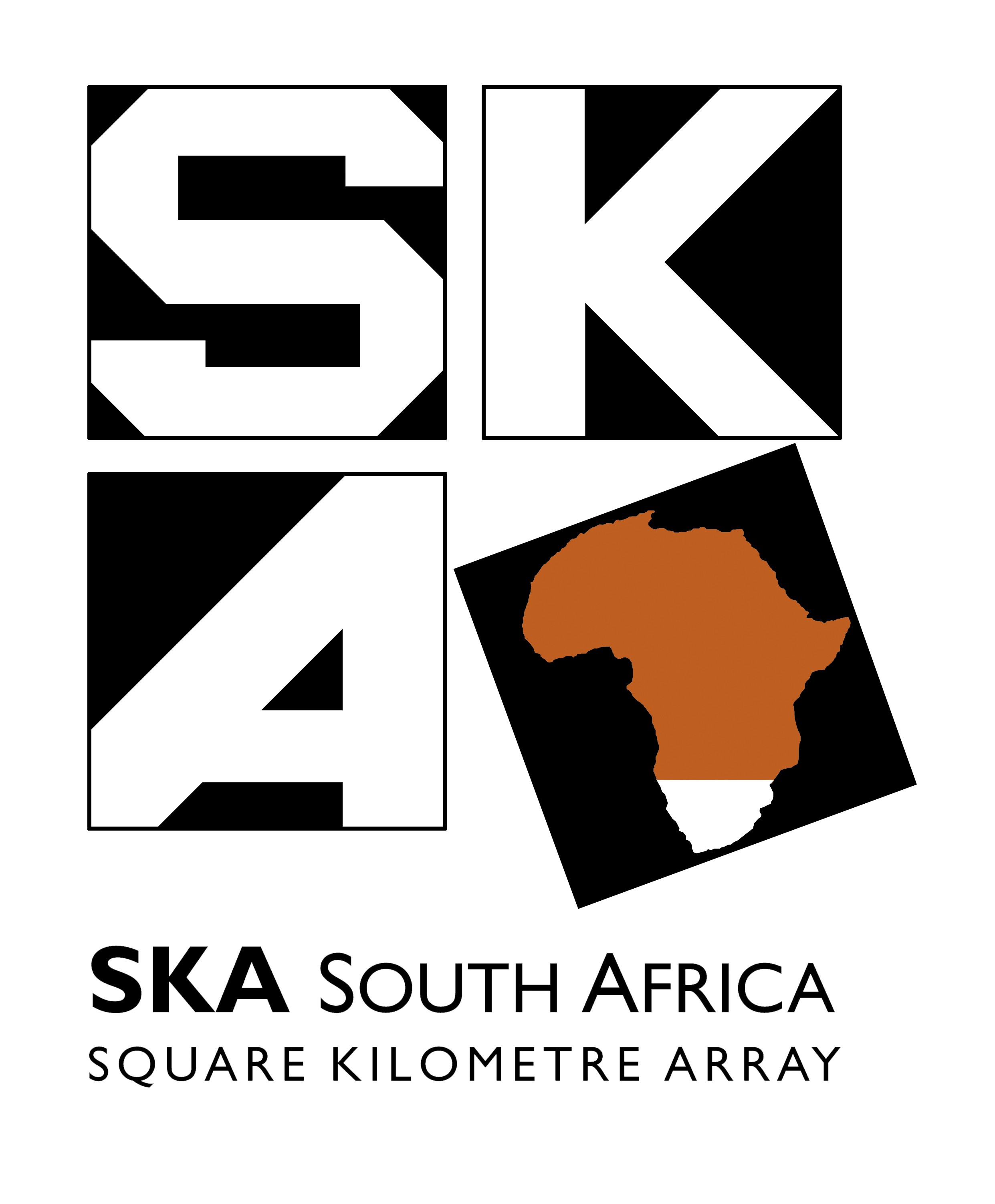}
\end{figure}

{\sf\Large
{\hfill An open invitation to the Astronomical Community\hfill}

{\hfill to propose Key Project Science with the South African\hfill}

\hfill{Square Kilometre Array Precursor \hfill}
\medskip\smallskip

{\centerline{\Huge\bf  MeerKAT}}

}

\bigskip
\bigskip

{\sc\bf R.S.\ Booth}

{\footnotesize\it Hartebeesthoek Radio Astronomy Observatory,
P.O.Box 443, Krugersdorp 1740, South Africa}\\
\indent{\footnotesize\it email: roy@hartrao.ac.za}

{\sc\bf W.J.G.\ de Blok}

{\footnotesize\it Department of Astronomy, University of Cape Town, Rondebosch 7700, South Africa.}\\
\indent{\footnotesize\it email: edeblok@ast.uct.ac.za}

{\sc\bf J.L.\ Jonas}

{\footnotesize\it Rhodes University, Dept.\ Physics \& Electronics, PO Box 94, Grahamstown 6410,
South Africa}\\
\indent{\footnotesize\it email: j.jonas@ru.ac.za}

{\sc\bf B.\ Fanaroff}

{\footnotesize\it SKA South Africa Project Office, 17 Baker St, Rosebank, Johannesburg, South Africa}\\
\indent{\footnotesize\it email: bfanaroff@fanaroff.co.za}

\medskip


\bigskip 
\emph{\large\underline{Proposal Submission deadline: March 15, 2010}}

\section{Introduction}
As possible hosts of the Square Kilometre Array (SKA), South Africa
and Australia are building SKA Precursor arrays: MeerKAT and ASKAP,
respectively. The two telescopes will complement each other well:
ASKAP will have a wider field of view but a smaller frequency range
and lower sensitivity, while MeerKAT will be more sensitive, have a
larger frequency range, but with a smaller field of view.  MeerKAT
will have additional shorter and longer baselines, giving it enhanced
surface brightness sensitivity as well as astrometric capability. It
is also envisaged that MeerKAT will have the capability of phasing-up
array elements and will, from time to time, participate in
the European, Australian and global VLBI networks.  This document gives a
short overview of the expected scientific capabilities as well as the
technical specifications of the MeerKAT telescope and invites the
community to submit Large Project proposals that take
advantage of the unique capabilities of the instrument.  In this
document we give a short description of the MeerKAT array in
Sect.~2. A brief summary of MeerKAT key science is given in Sect.~3.
More detailed technical specifications and the array configuration are
given in Sect.~4, with the proposal format and policies listed in
Sect.~5. A summary is given in Sect.~6.

\section{MeerKAT}
The Karoo Array Telescope MeerKAT will be the most sensitive
centimetre wavelength instrument in the Southern Hemisphere; it will
provide high-dynamic range and high-fidelity imaging over almost an
order of magnitude in resolution ($\sim 1$ arcsec to $\sim 1$ arcmin at 1420
MHz). The array will be optimized for deep and high fidelity imaging
of extended low-brightness emission, the detection of micro-Jansky
radio sources, the measurement of polarization, and the monitoring of
radio transient sources.  It will be ideal for extragalactic HI
science, with the possibility of detecting extremely low column
density gas, but high resolution observations of individual galaxies
are also possible. Its sensitivity, combined with excellent
polarisation purity, will also make it well suited for studies of
magnetic fields and their evolution, while its time domain capability
will be ideal for studying transient events. Planned high frequency
capabilities will give access to Galactic Centre pulsars, and make
possible measurements of CO in the early Universe at redshifts $z\sim
7$ or more.  

MeerKAT is being built in the Karoo, a part of South Africa's Northern
Cape region which has a particularly low population density. Part of
the Northern Cape, through an Act of Parliament, is being declared a
Radio Astronomy Reserve. The approximate geographical coordinates of
the array are longitude 21$^{\circ}$23$'$E and latitude
30$^{\circ}$42$'$S.  MeerKAT will be an array of 80 antennas of 12 m
diameter, mostly in a compact two-dimensional configuration with 70\%
of the dishes within a diameter of 1 km and the rest in a more
extended two-dimensional distribution out to baselines of 8 km. An
additional seven antennas will be placed further out, giving E--W
baselines out to about 60 km. These will give a sub-arcsecond
astrometric capability for position measurements of detected sources
and enable their cross-identification with other instruments. The
extra resolution will also drive down the confusion limit for
surveys. Finally, it will be possible to phase the central core as a
single dish for VLBI observations with the European and Australian
networks.  The initial frequency range of the instrument, in 2013,
will be from 900 MHz to approximately 1.75 GHz. This will be extended
with a 8--14.5 GHz high frequency mode in 2014. The lower frequency range
will be further extended to 580 MHz--2.5 GHz in 2016.

\section{Science with MeerKAT}
We envisage a range of scientific projects for which MeerKAT will have
unique capabilities. These include extremely sensitive studies of
neutral hydrogen in emission --- possibly out to $z = 1.4$ using
stacking and gravitational lens amplification --- and highly
sensitive continuum surveys to $\mu$Jy levels, at frequencies as low
as 580~MHz. The good polarisation properties will also enable
sensitive studies of magnetic fields and Faraday rotation to be
conducted. MeerKAT will be capable of sensitive measurements of
pulsars and transient sources. The high frequency capability will
facilitate such measurements even towards the centre of the
Galaxy. MeerKAT will be sensitive enough to conduct molecular line
surveys over a wide frequency range: not only will Galactic Surveys of
hydroxyl and methanol masers be possible, but at longer wavelengths
(pre-biotic) molecules can also be detected. At the highest
frequencies, CO at $z > 7$ may be detectable in its $J=1-0$ ground
state transition.

Many of the applications of the Precursor instruments are driven by
the SKA scientific programme, which has been described in a special
volume of ``New Astronomy Reviews'' (vol.\ 48, 2004) as well as in the
description of the SKA Precursor ASKAP science in volume 22 of
``Experimental Astronomy'' (2008). We do not intend to repeat the full
scientific motivation here, but present a brief outline of the
particular scientific programmes in which we believe MeerKAT will
excel, and which we hope will excite collaborations from among
astronomers world-wide. The location and science goals of MeerKAT lend
themselves to intensive collaborations and joint projects with the
many facilities at other wavelengths available in the southern
hemisphere. Combinations with large mm-arrays like ALMA, but also
SALT, VISTA, VST, APEX, VLT and Gemini South, to name but a few of the
many instruments available, should prove to be fruitful.

\subsection{Low frequency bands (580 MHz - 2.5 GHz)}
\subsubsection{Extragalactic HI science and the evolution of galaxies}
Deep HI observations are a prime science objective for MeerKAT. In the
general SKA Precursor environment, initial indications are that
MeerKAT will be the pre-eminent southern hemisphere HI observation
facility for regions $\sim 10$ deg$^2$ or less and for individually
significant HI detections out $z \sim 0.4$. For surveys of $\sim 30$
deg$^2$ or more, ASKAP will likely be the instrument of choice. Where
exactly the ideal balance point lies between these facilities will
continue to evolve as our understanding of both telescopes and their
survey capabilities improve. Together, these facilities offer the
opportunity to create a comprehensive tiered HI program covering all
epochs to redshift unity and beyond.

\paragraph{Deep HI surveys}
The formation of stars and galaxies since the epoch of re-ionisation
is one of today's fundamental astrophysical problems. Determining the
evolution of the baryons and the dark matter therefore forms one of
the basic motivations for the SKA and MeerKAT.  A one-year deep HI
survey with MeerKAT would give direct detections of HI in emission out
to $z \sim 0.4$, and using the stacking technique and gravitational
lensing would enable statistical measurements of the total amount of
HI out to even higher redshifts up to $z \sim 1.4$. The advantage of
the stacking technique is that high signal-to-noise detections of
individual galaxies are not necessarily required.  Using previously
obtained (optical and near-IR) redshifts, one can shift even very low
signal-to-noise spectra (which would not on their own constitute a
reasonable detection) such that all the spectral lines fall into a
common channel and then stack the spectra to produce an average
spectrum. Since spectroscopic redshifts are required, the HI survey
will need to overlap with an existing or near-future redshift survey
field. A further sensitivity enhancement involving gravitational lens
amplification may be exploited in appropriate fields.

\paragraph{Studies of the Low Column Density Universe}
Galaxies are believed to be embedded in a ``cosmic web'', a
three-dimensional large scale structure of filaments containing the
galaxy groups and clusters. It is now reasonably certain that most of
the baryons do not, in fact, reside in galaxies, but are found outside
galaxies spread along this ``web''. The material is, however, tenuous
and the neutral fraction is small. It has possibly been seen in a few
lines of sight as absorption features against background sources but a
direct detection of the cosmic web would significantly improve our
understanding of the baryon content of the universe. The cosmic web
may be the source of the HI seen around galaxies taking part in the
so-called cold accretion process. The material is expected to have
column densities around $10^{17-18}$ cm$^{-2}$. Surveys for this low
column density HI would likely be conducted by targeting a number of
nearby galaxies. Assuming a 20 km s$^{-1}$ channel spacing (the
expected FWHM line-width of an HI line), one would need to integrate
with MeerKAT for about 150 hours for a $5\sigma$ detection of a
$10^{18}$ cm$^{-2}$ signal at a resolution of $\sim 90''$. Assuming only
night-time observing, this means that a direct detection of the low
column density gas around galaxies can be done for a different galaxy
every two weeks, thus rapidly enabling comparisons of morphology and
properties of the low column density gas for a wide range in Hubble
type.  Depending on the flexibility of the correlator and the presence
of background sources these observations could also be used to probe
the low column density universe at higher redshifts using HI
absorption.

\paragraph{A high-resolution survey of the HI distribution in 1000 nearby galaxies}
Detailed, high-resolution (sub-kpc) observations of the interstellar
medium in nearby galaxies are crucial for understanding the internal
dynamics of galaxies as well as the conversion from gas into
stars. Recent high-resolution HI surveys (such as The HI Nearby Galaxy
Survey THINGS performed at the VLA) clearly showed the power of
obtaining detailed 21-cm observations and combining them with
multi-wavelength (particularly infrared and UV) data to probe galaxy
evolution and physical processes in the interstellar medium. A more
extensive sensitive high-resolution survey in the southern hemisphere
will provide important data on star formation and dark matter in a
large range of galaxy types in a wide range of environments.  A single
8-hour observation with MeerKAT rivals the THINGS VLA observations in
terms of resolution and column density sensitivity.  This is
particularly relevant with the advent of sensitive surveys and
observations of the molecular and dust component of the ISM by
Herschel and, in the future, ALMA. These combined studies will provide
the local calibration point against which higher redshift studies can
be gauged. The presence of major optical, IR and sub-mm telescopes in
the southern hemisphere make such a multi-wavelength approach
desirable.

\paragraph{An HI absorption line survey (and OH mega-masers)}
Most HI absorption measurements have been made at optical wavelengths
in damped Lyman-$\alpha$ systems. Such systems are prone to biases, as
from the ground it is only possible to observe the line red-shifted to
$z \simeq 1.7$. Furthermore, dust obscuration probably causes the
observations to be biased against systems with a high
metallicity. Such biases are not a problem for the HI line. As radio
continuum sources span a large range of redshift, MeerKAT observations
should detect absorption over the low frequency band to $z = 1.4$. The
VLBI capability of the array should enable high-resolution follow up
with either the EVN or the Australian array, depending on declination
and redshift.  A judicious choice of frequency bands for the HI
absorption line survey will also pick up narrow band emission from
hydroxyl, OH. The extragalactic OH emission, especially at 1667 MHz,
will delineate mega-masers, maser emission associated mainly with
interacting or starburst galaxies, some of which will show
polarisation (Zeeman-)patterns from which line of sight magnetic
fields may be inferred.

\subsubsection{Continuum measurements}
\paragraph{Ultra-deep, narrow-field continuum surveys with full polarisation measurements}
The MeerKAT-ASKAP complementarity discussed in relation to deep HI
surveys applies also to surveys in radio continuum and
polarisation. ASKAP will survey the entire southern sky to an rms
noise limit of 50 $\mu$Jy per beam in 1 year of observing time, while
we envisage that MeerKAT will, in the first instance, make a number of
deep pointings in fields that are already being studied at other
wavelengths (e.g., Herschel ATLAS, Herschel HerMES, SXDS, GOODS and
COSMOS). Within the 1 deg$^{2}$ MeerKAT field at 1400 MHz, a
conservative $(5\sigma)$ estimate of the sensitivity is 7 $\mu$Jy per
beam in 24 hours with 500 MHz bandwidth. This scales to 0.7 $\mu$Jy in
100 days. Dealing with confusion at this level will require judicious
use of the long baselines (using the 60 km E--W spur).  This exciting
work will study radio-galaxy evolution, the AGN-starburst galaxy
populations and their relationship, perhaps through AGN feedback, in
unprecedented detail, so addressing the evolution of black holes with
cosmic time. It may even reveal a new population of radio sources and
address the enigmatic far-IR-radio correlation at high redshifts.

\paragraph{Magnetic Fields}

Polarisation studies will, in the first instance, use the full low
frequency band in determining rotation measures for the fraction of stronger
sources. The intra-cluster medium has been shown to be
magnetised and it would seem that magnetic fields play a critical role
in the formation and evolution of clusters. A rotation measure survey
of several clusters would be feasible with MeerKAT through
observations, in several low frequency bands, of sources within and
behind the cluster.

\paragraph{Galactic studies and the Magellanic Clouds}

As well as important extragalactic science, we envisage much interest
in Galactic and Magellanic surveys with MeerKAT, both in HI, for
measurements of dynamics, together with measurements of the Zeeman
effect and determinations of the line of sight component of the
magnetic field. Similar measurements may be made in the 4 ground state
OH lines.  There are relatively few measurements of all 4 ground state
OH lines.  MeerKAT's wide band and high spectral resolution will
enable such measurements, which give valuable information on the
excitation temperature and the column density of the cold gas
component.  Very few know interstellar molecules have lines in bands
below L-band. An interesting exception is methanol whose transition at
830 MHz was the first to be measured and so identify the
species. While a Galactic survey in this line would be interesting, it
will also be exciting to perform a census of low frequency transitions
in directions towards the Galactic Centre and Sgr B. Molecules with
transitions at the lower frequencies tend to be bigger and could be
more important as pre-biological molecules and their potential
importance for the origin of life.

\subsection{High frequency science (8 - 14.5 GHz)} 
\subsubsection{Pulsars and transients} 

The high-frequency capability of MeerKAT will be particularly useful
for studies of the inner Galaxy.  We expect the population of pulsars
in the Galactic Centre to be large, but it is being obscured by
interstellar scattering, which cannot be removed by instrumental
means. Observing at sufficiently high frequencies ($\sim$ 10 GHz or
higher) the pulsar population can be revealed. A number of these
pulsars will be orbiting the central supermassive black hole. The
orbital motion of these pulsars will be affected by the spin and
quadrupole moment of the black hole. By measuring the effects of
classical and relativistic spin-orbit coupling on the pulsar's orbital
motion in terms of precession, traced with pulsar timing, we can test
the cosmic censorship conjecture and the no-hair theorem. The
technical requirements in terms of data acquisition and software
involved in this application are, however, challenging and a large
community effort will be needed to be successful in this challenging
but exciting science goal.

\subsubsection{High-$z$ CO}

While HI emission has been difficult to detect at even moderate
redshifts, CO has been detected with the VLA in the $J = 3-2$
rotational transition at $z = 6.4$. The new large millimetre array,
ALMA, will detect higher CO rotational transitions at $z > 6$, and it
may be instructive to measure the ground state rotational transition
for comparison. The new EVLA will open up CO$(1-0)$ surveying,
particularly in the northern sky. In the southern sky, MeerKAT will be
its counterpart, but with a larger field of view and sky
coverage. MeerKAT at 14.5 GHz will facilitate the detection of CO$(1-0)$
emission at $z > 6.7$, and the ground state transition of HCO$^+$ at
$z > 4.9$. It will be important to exploit such commensality with
ALMA, and to compare the atomic and molecular content of galaxies as a
function of redshift, since recent studies show that the molecular
hydrogen proportion may increase with redshift.

\subsection{VLBI science}

The availability of the phased central MeerKAT antennas, as the
equivalent of an $\sim 85$ m diameter single-dish antenna, will have a
profound effect on the highest resolution measurements, made with
VLBI.  A phased MeerKAT will both increase the $uv$-coverage available
in the South, and provide great sensitivity on the longest baselines,
where visibility amplitude is often low as sources are becoming
resolved.  This and the recently demonstrated e-VLBI capacity with the
Hartebeesthoek Radio Telescope will create great demand for the phased
MeerKAT in the VLBI networks.  A particular application of
significance with the European VLBI network will be wide field imaging
VLBI of sources in Deep Fields. High sensitivity VLBI studies of the
Hubble Deep Field revealed $\mu$Jy sources, many of which were
starburst galaxies. In one case a radio-loud AGN was detected in a
dust obscured, $z = 4.4$ starburst system, suggesting that at least
some fraction of the optically faint radio source population harbour
hidden AGN.

Wide-field imaging developments with the EVN and MeerKAT will not only
produce more interesting fine detail on galaxies in the early
universe, they will also be a test bed for the SKA. Furthermore, the
presence of the HESS high-energy telescope and its successor in close
proximity to South Africa will enhance the importance of the southern
VLBI arrays for studying the radio component of high-energy gamma ray
sources.  

Another field of interest for VLBI arrays including MeerKAT is that of
(narrow band) masers.  Trigonometric parallaxes of maser spots are
refining distance measurements in the Milky Way and improving our
knowledge of the structure and dynamics of the Galaxy.  Both hydroxyl
(1.6 GHz) and methanol (12 GHz lines) still reveal new and interesting
properties, like alignments and discs, in regions of star
formation. Studies of OH masers as well as the radio continuum in
starburst galaxies like Arp 220 are revealing strings of supernovae
and strange point sources whose spectra have high a frequency
turnover, and might even be indicative of 'hypernovae'.

\subsection{Other Science}
We have described some of the exciting science that will be done with
MeerKAT, but the new instrument will have the potential to do much
more. All-sky surveys at 600 MHz and 8 GHz are possible, as well as
Galactic polarisation measurements and deep studies of magnetic
fields, and science requiring high brightness sensitivity at high
frequency (e.g., of the Sunyaev-Zel'dovich effect). There are
possibilities for pulsar surveys and much more. While initially
MeerKAT science will focus on the key science areas described here, we
also welcome inventive proposals beyond the ones suggested in this
document that make use of unique scientific capabilities of MeerKAT.

\section{MeerKAT: specifications and configuration}

MeerKAT will consist of 80 dishes of 12 m each, and it will be capable
of high-resolution and high fidelity imaging over a wide range in
frequency. The minimum baseline will be 20~m, the maximum 8 km. An
additional spur of 7 dishes will be added later to provide longer
(8--60 km) baselines. It is intended that the final array will have 2
frequency ranges: 0.58--2.5 GHz and 8--14.5 GHz, with the full frequency
range gradually phased in during the first years of the array.

MeerKAT commissioning will take place in 2012 with the array coming
online for science operations in 2013. Table 1 summarizes the final
MeerKAT specifications. Table 2 gives an overview of the various
phases of the MeerKAT construction and commissioning leading up to
these final specifications.  

MeerKAT will be preceded by a smaller prototype array of seven
antennas, called KAT-7. The commissioning of this science and
engineering prototype will start in 2010, with test science
observations expected later that year. KAT-7 will be used as a test
bed for MeerKAT, as well as for the data reduction pipelines etc. and
is more limited in its science scope, with smaller frequency coverage
(1.2-1.95 GHz), and longest and shortest baselines of 200m and 20m
respectively, as also indicated in Table 2.

The \emph{maximum} processed bandwidth on MeerKAT will initially be
850 MHz per polarization.  This will gradually be increased to 4 GHz.
There will be a fixed number of channels, initially 16384, though this
may be increased if demand requires.  By choosing smaller processed
bandwidths (down to 8 MHz) the velocity resolution may be increased.
Note that not all combinations of specifications will be realized. For
example, fully correlating the long 60 km baselines with the full
array at the minimum sample time with the maximum number of channels
will result in prohibitive data rates.

\begin{table}
\begin{center}
\caption{MeerKAT final system properties}
\begin{tabular*}{0.76\textwidth}{@{\extracolsep{\fill}}|l|l|}
\hline
  Number of dishes$^a$ & 80 (central array)\\
  & + 7 (spur)\\
  Dish diameter &    12 m \\
  Aperture efficiency &  0.7\\
  System temperature & 30 K\\
  Low frequency range$^a$ & 0.58--2.5 GHz\\
  High frequency range$^a$ & 8--14.5 GHz\\
  Field of view & 1 deg$^2$ at 1.4 GHz \\
                & 6 deg$^2$ at 580 MHz \\
                & 0.5 deg$^2$ at 2 GHz \\
  $A_e/T_{\rm sys}$       & 200 m$^2$/K\\
Continuum imaging dynamic range$^b$ & 1:10$^5$\\
Spectral dynamic range$^b$ & 1:10$^5$\\
Instrumental linear & \\
\ \ polarisation purity & $-25$ dB across field\\
Minimum and maximum bandwidth & \\
\ \ per polarization$^a$ & 8 MHz--4 GHz\\
Number of channels & 16384\\
Minimum sample time & 0.1 ms\\
Minimum baseline  & 20 m\\
Maximum baseline & 8 km (without spur)\\
                 & 60 km (with spur)\\
\hline
\end{tabular*}\\
\vspace{2pt}
\emph{Notes:} $a$: \emph{Final values. See Table 2 for roll-out schedule.}\\ $b$: \emph{Dynamic range defined as rms/maximum.}
\end{center}
\end{table}

\begin{table}
\begin{center}
\caption{MeerKAT Phasing Schedule}
\begin{tabular}{|l|r|r|r|r|r|}
\hline
                          &  KAT-7         & Phase 1      & Phase 2           & Phase 3        & Phase 4\\
                          &  2010          & 2013         & 2014              & 2015           & 2016   \\
\hline
Number of dishes          &  7             & 80           & 80                & 87             & 87     \\
Low freq.\ range (GHz)      &  1.2--1.95 & 0.9--1.75 & 0.9--1.75     & 0.9--1.75  & 0.58--2.5\\
High freq.\ range (GHz)     &  ---           & ---          & 8--14.5        & 8--14.5   & 8--14.5 \\
Maximum processed &  &  & & & \\ 
\ \ bandwidth (GHz)  &  0.256       & 0.850      & 2            &  2  &    4  \\
Min.\ baseline  (m)        &  20           & 20      &  20      &  20      &  20      \\
Max.\ baseline  (km)        &  0.2         & 8  & 8 & 60  & 60 \\
\hline
\end{tabular}
\end{center}
\end{table}

\subsection{Configuration}

The MeerKAT array will be constructed in multiple phases (see Table 2). The first
phase will consist of 80 dishes distributed over two components.
      
\begin{itemize} 
\item 1. A dense inner component containing 70\% of the dishes. These are 
distributed in a two-dimensional fashion with a Gaussian $uv$-distribution with a 
dispersion of 300 m, a shortest baseline of 20 m and a longest baseline of 1 km.
\item 2. An outer component containing 30\% of the dishes. These are also 
distributed resulting in a two-dimensional Gaussian $uv$-distribution with a 
dispersion of 2500 m and a longest baseline of 8 km.
\end{itemize}
This will be followed by a second phase which will involve the
addition of a number of longer baselines.
\begin{itemize}
\item 3. A spur of an additional 7 antennas will be distributed along
  the road from the MeerKAT site to the Klerefontein support base,
  approximately 90 km SE from the site. This will result in E--W
  baselines of up to 60 km.  The positions of these antennas will be
  chosen to optimize the high-resolution performance of the array to
  enable deep continuum imagine and source localisation.  
\end{itemize}

Figure 1 shows a concept configuration of components 1 and 2 listed
above.  Positions of individual antennas may still change pending
completion of geological measurements, but will remain consistent with
the concept of a 70/30 division between a 1 km maximum baseline core
and an 8 km maximum baseline outer component. X and Y positions of the
antennas with respect to the centre of the array are given in the
Appendix.  Representative $uv$-distributions for observations of
different duration towards a declination of $-30^{\circ}$ are given in
Fig.\ 2. A histogram of the total baseline distribution for an 8h
observation towards $-30^{\circ}$ is given in Fig.~3.

\begin{figure}[t]
\centering
\includegraphics[width=\textwidth]{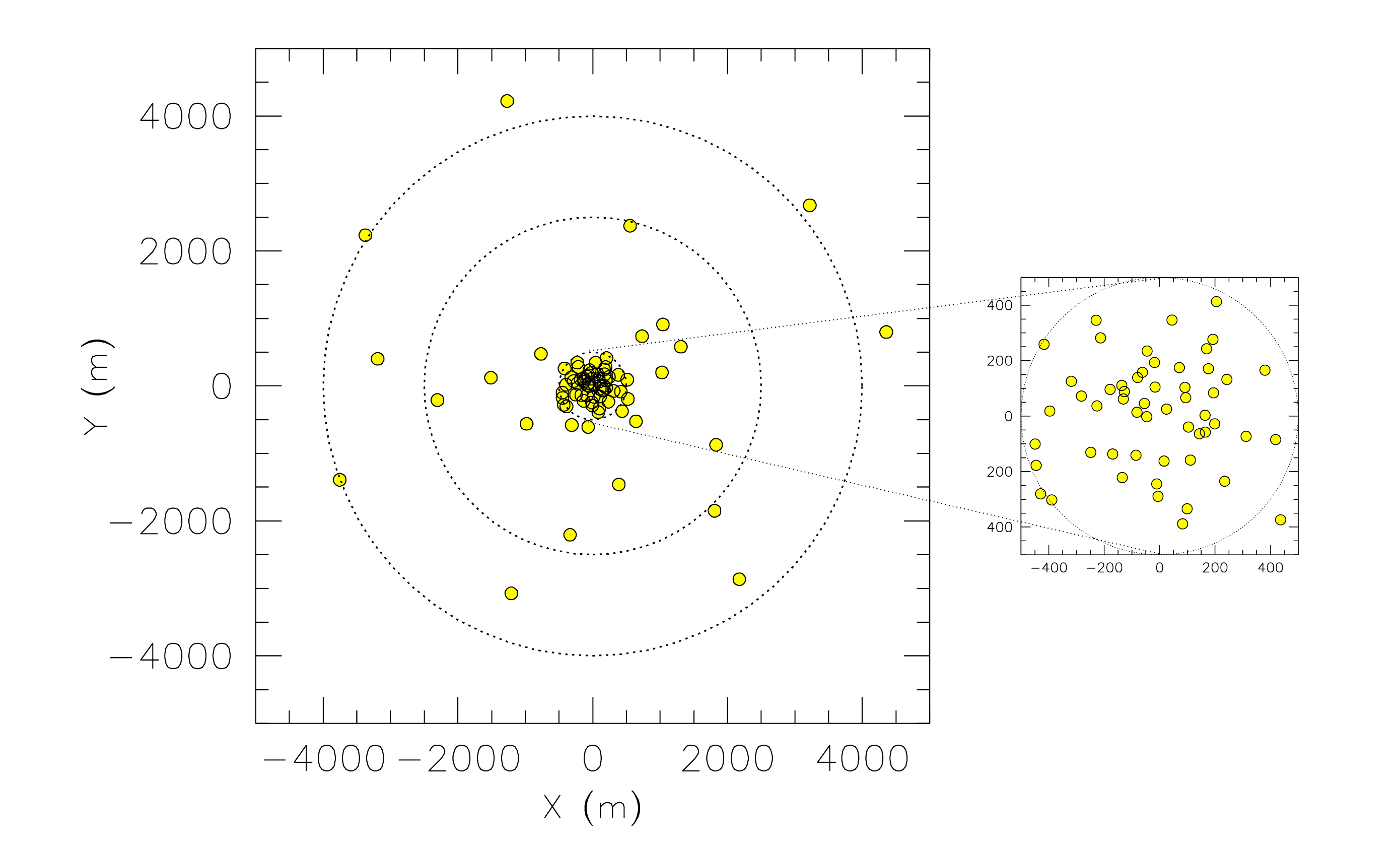}
\caption{\emph{Overview of the MeerKAT configuration. The inner component contains 70\% of the dishes, 
using a two-dimensional Gaussian uv-distribution with a dispersion of 300 m and a longest baseline 
of 1 km. The outer component contains 30\% of the dishes, and is distributed as a two-dimensional 
Gaussian uv-distribution with a dispersion of 2.5 km and a longest baseline of 8 km. The shortest 
baseline is 20 m. The three circles have diameters of 1, 5 and 8 km. The inset on the right shows a 
more detailed view of the inner core.}}
\end{figure}

\begin{figure}
\centering
\includegraphics[width=0.495\textwidth]{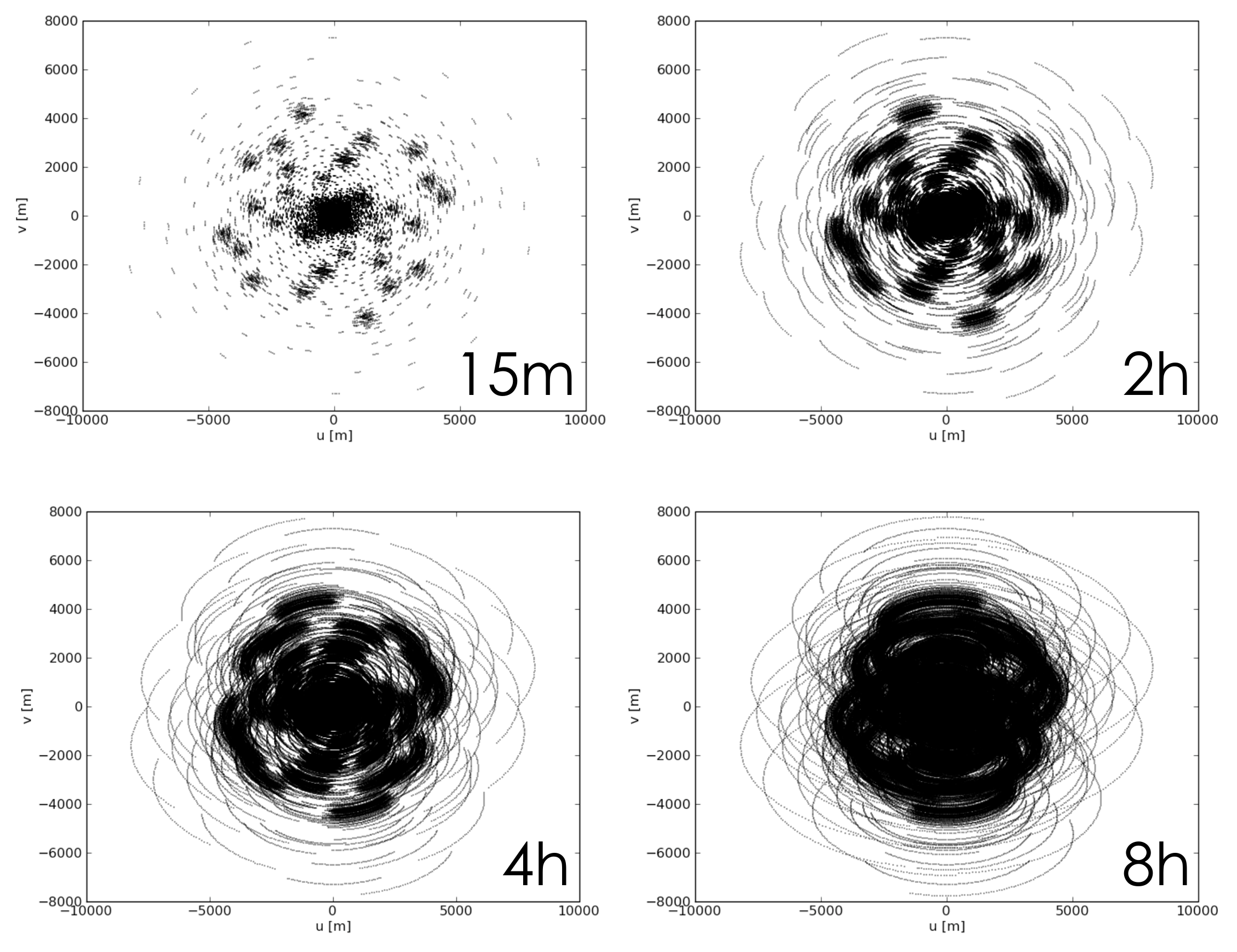}
\includegraphics[width=0.45\textwidth]{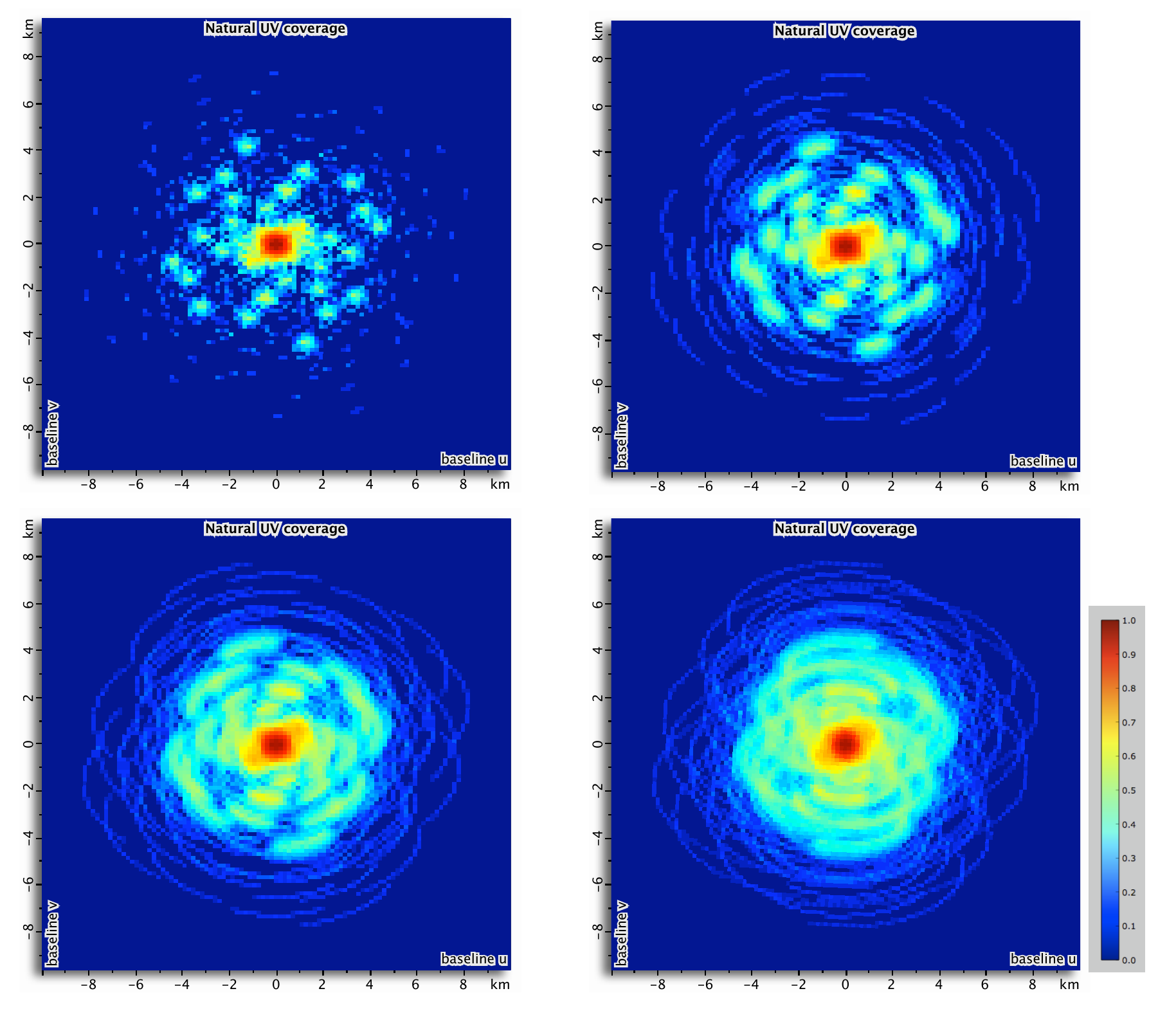}
\caption{\emph{Left panel: uv distribution of the MeerKAT array for observations towards declination $-30^{\circ}$,
with the observing time indicated in the sub-panels. Right panel: density of uv-samples for the 
corresponding observations in the left panel.}}
\end{figure}

\begin{figure}
\centering
\includegraphics[width=0.7\textwidth]{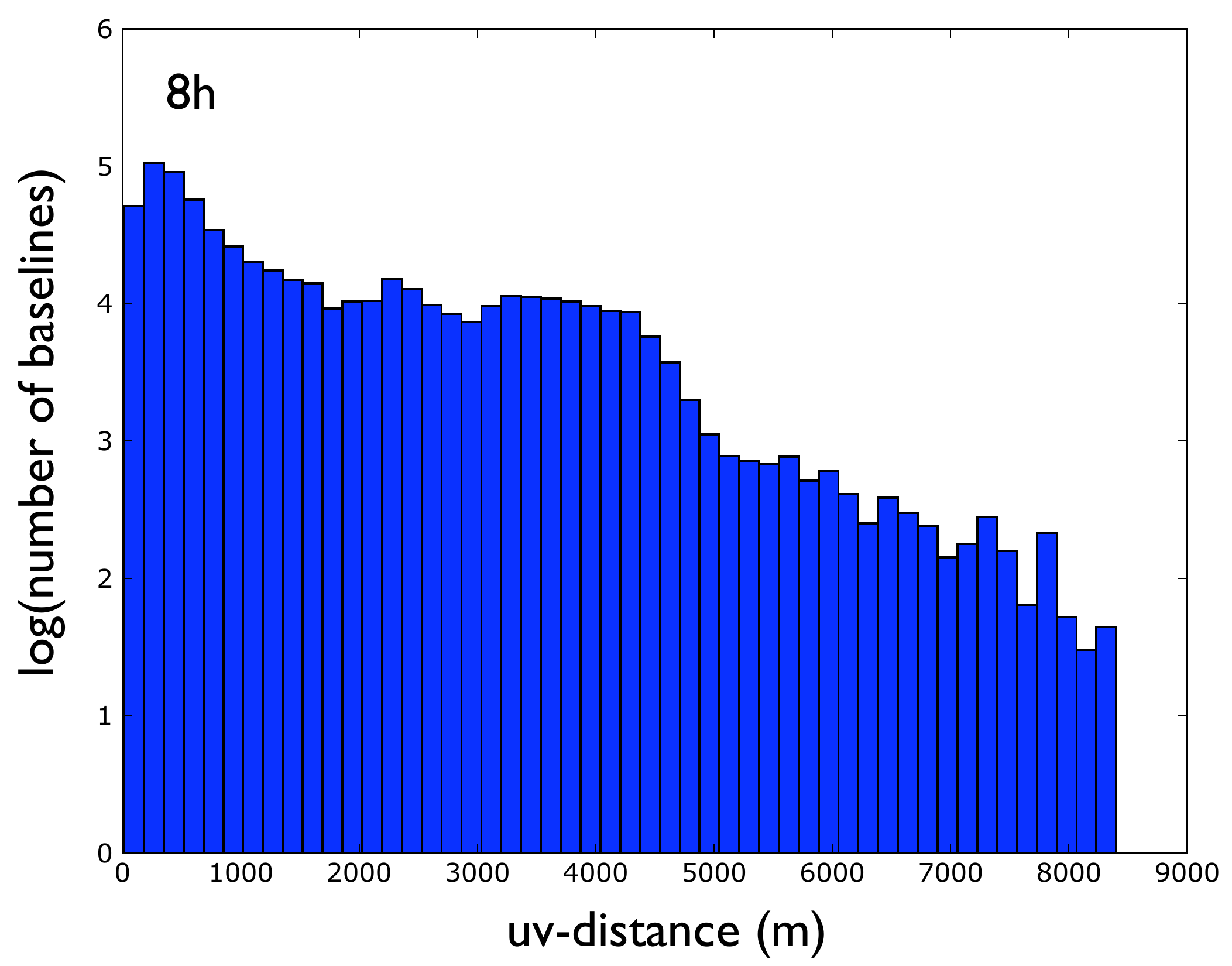}
\caption{\emph{Histogram of the uv-distance for an 8h observation towards -30$^{\circ}$. The histogram numbers 
assume 5 min sample integration times.}}
\end{figure}

\begin{figure}
\centering
\includegraphics[width=0.7\textwidth]{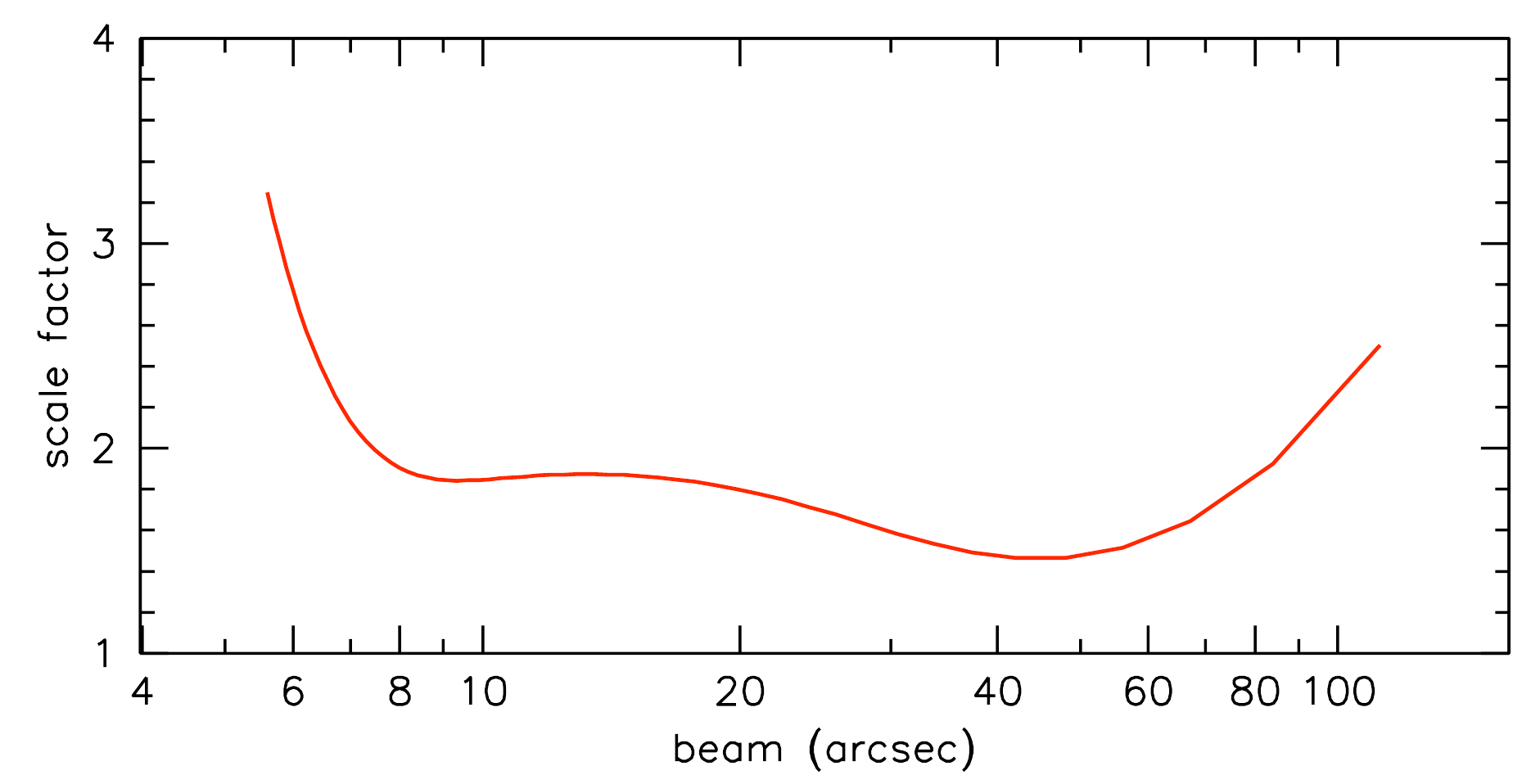}
\caption{\emph{Scale factor by which the naturally-weighted noise will be increased
when using tapering to obtain resolution shown on the horizontal axis,
assuming a frequency of 1420 MHz.}}
\end{figure}

\subsection{Sensitivity}

The multi-resolution configuration of the telescope means a taper or
similar kind of weighting of the $uv$-samples needs to be used to
produce a synthesized beam of the desired resolution. Tapering will
increase the noise with respect to natural weighting (as it reduces
the effective number of $uv$ samples). Figure 4 and Table 3 list the
correction factors to be applied to the expected noise for tapering to
a desired resolution with respect to the untapered, naturally-weighted
noise, assuming a frequency of 1420 MHz.

As a guideline, for an 8 hr spectral line observation, assuming a
channel width of 5 km s$^{-1}$ (23.5 kHz at 1420 MHz), a system
temperature of 30K at 1420 MHz, two polarizations, and a system
efficiency of 0.7, the expected untapered, naturally-weighted
5$\sigma$-noise level is 1.8 mJy/beam. Similarly, a 24 hour continuum
observation with a similar setup, but a 500 MHz bandwidth gives a
5$\sigma$ naturally weighted noise level of 7.2 $\mu$Jy. These
are naturally-weighted, untapered noises and need to be scaled with the
factors listed in Table 3 and Figure 4 for the desired resolution.

\begin{table}
\begin{center}
\caption{Scale factors as shown in Fig.\ 4}
\begin{tabular}{|c|c|}
\hline
Beam size at & Scale-\\
1420 MHz     & factor\\
(arcsec) & \\
\hline
6 & 2.7 \\
8 & 1.9 \\
10 &1.8\\
20 &1.8\\
40 &1.5\\
60 &1.6\\
80 &1.9\\
100 &2.2\\
\hline
\end{tabular}\\
\vspace{2pt}
\end{center}
\end{table}

\section{Call for Large Project proposals}

\subsection{Key Science Areas}

MeerKAT observing time will be allocated for Large Project proposals
and shorter PI proposals, with the intention that 75\% of telescope time
will be made available for the Large Projects during the first 5
years. The current call for proposals only applies to the Large
Projects. A separate call for short proposals will be made at a later
stage. As described in Section 2, the MeerKAT Large Projects are
envisaged to cover the following Key Science areas, although well-motivated
proposals in other appropriate fields are welcome:
\begin{itemize}
\item \emph{Neutral hydrogen}
  \begin{itemize}
  \item Deep emission and absorption studies out to redshift $z = 1.4$.
  \item Deep observations of selected groups or galaxies and detection
    of the cosmic web.
  \item A high-resolution survey of a substantial number of galaxies
    in the nearby universe, coordinated with surveys at other
    wavelengths.  
  \item Targeted observations of selected regions of the Galaxy and
    the Magellanic Clouds in HI and OH.
  \end{itemize}
\item \emph{Deep continuum observations}
  \begin{itemize}
  \item Deep observations, with polarization in selected regions.
  \item Observations of polarized emission in cluster magnetic fields with a goal 
of understanding the role of magnetism in the Universe.
  \end{itemize}
\item \emph{Pulsars and transients}
  \begin{itemize}
  \item High frequency pulsar searches and monitoring in regions towards the 
Galactic centre.
  \item Long duration ``staring'' observations with high time resolution with 
transient detection as a goal.
  \end{itemize}
\item \emph{Deep field searches for high-z CO and HCO$^+$}
  \begin{itemize}
  \item Searches for high-$z$ emission from the ground state ($J=1-0$)
    CO at 14.5 GHz in selected regions, perhaps in concert with ALMA
    observations.
  \end{itemize}
\end{itemize}

\subsection{Large Project proposals}

We invite applications for Large Projects from the South African,
African and international astronomical community.  A Large Project is
defined as a project with duration $\geq 1000$ hours, which addresses
one of the Key Science Areas listed above in a coherent manner,
utilizing the resolution and/or sensitivity strengths of MeerKAT (thus
includes Projects with a large complementarity with other SKA
Precursor or Pathfinder projects). Large Projects should aim to
achieve their goals in a manner not possible by a series of small
projects (for which a separate call will be issued at a later stage).

This call for proposals is open to South-African, African and
international research teams.  We encourage including South African
collaborators because of their familiarity with the instrument and to
facilitate communication with the engineering and commissioning
teams. There will be a limited number of projects per Key Science area
and proposers are therefore encouraged to coordinate and collaborate
prior to submission. The final decision on allocations of observing
time will be through an international peer-review process and
successful teams will be invited to further develop their projects by
mid-2010.

Initially, the MeerKAT operations team will provide calibrated
$uv$-data, however, as experience with the array grows, development
of other data products will be undertaken by the MeerKAT science and
engineering team, in collaboration with the research teams.

Further data reduction pipelines, simulations and preparations will be
the responsibility of the research teams.  We envisage grants being
available for team members to spend time in South Africa to become
familiar with the MeerKAT instrument and its software and for training
local students and postdocs who want to be part of the proposal. We
also plan on making available bursaries for postdocs and students
(preferably based in South Africa) to support the projects. We will
also consider supporting team meetings. Teams will have reasonable
access to advice and support from the MeerKAT team when operations
commence.

Teams are asked to specify in their proposals what their final and
intermediate data products and deliverables will be. We also request
they indicate how their project will make optimal use of the MeerKAT
roll-out phases as specified in Table~2. We encourage teams to
indicate their publication policy and to also indicate the
possibilities for early results.  There will be an 18 month
proprietary period on any data, counted from when the observation was
made. Teams can apply to the MeerKAT director for an extension, 
which will be given only if justified by the nature of the
project and if there is evidence of progress towards
publication of substantive results.

\subsection{Proposal format}

Large Project proposals must be in PDF format, with a minimum font
size of 11 pt and reasonable page margins. Proposals must contain the
following chapters:

\begin{enumerate}
\item	Abstract (maximum of half a page).
\item	A list of team members and their roles and affiliations (2 pages maximum).
\item	Scientific case --- a comprehensive discussion of the astrophysical importance of 
the work proposed (6 pages maximum).
\item Observational strategy --- the most effective way to perform the
  proposed observations. Teams should indicate milestones and
  intermediate results and data products (2 pages maximum).
\item Complementarity --- coordination with surveys at other
  wavelengths or with other SKA Precursor or Pathfinder instruments (2 page maximum).
\item Strategy for data storage and analysis (2 pages maximum).
\item Organisation --- specify team leaders, their roles and their
  sub-groups, with a clear overview of how the team will divide the
  identified tasks (2 pages maximum).
\item The publication strategy --- specify intermediate and final data releases
  and availability (1 page maximum).
\item Team budget --- teams should indicate and justify budgetary
  needs that cannot be met as part of established observing and data
  analysis procedures commonly in use at other observatories and research
  institutes (1 page maximum).
\item Requirements for special software --- teams are asked to indicate
  special computing needs. These must be communicated to
  the project as early as possible and the project team will be
  expected to provide as much of that specialized software as possible
  into the MeerKAT observing system. However, that effort will be
  considered positively when awarding time to the proposal (1 page maximum).
\item	Public outreach and popularization efforts (1 page maximum).
\end{enumerate}

\subsection{Proposal submission}

The deadline for submission of proposals is  {\bf March 15, 2010}.
Proposals must be submitted as a PDF file. Proposal submission procedures will
be available on the MeerKAT website {\tt http://www.ska.ac.za/meerkat} from about one
month before the deadline for submission.

An evaluation of proposals will be made in mid-2010 by an
international peer review committee and successful teams will be
invited to further develop their proposals.

For questions regarding MeerKAT technical specifications and proposal
preparation please contact Justin Jonas, {\tt j.jonas@ru.ac.za}.

\section{Concluding remarks}

With this short discussion of the MeerKAT scientific goals and
applications, we invite Large Proposals from teams
prepared to help design observational programmes, and help contribute
to developing simulations, and advanced observing methods and analysis
techniques.

MeerKAT will be capable of very exciting science.  It will be a major
pathfinder to the SKA, giving insights into many of the technical
challenges of the SKA, but also giving a glimpse of the new
fundamental studies that the SKA will facilitate.

\bigskip \emph{We acknowledge the help and stimulation given by
  members of the MeerKAT International Scientific Advisory Group:
  Bruce Bassett, Mike Garrett, Michael Kramer, Robert Laing, Scott
  Ransom, Steve Rawlings and Lister Staveley-Smith. We also thank
  Michael Bietenholz, Bradley Frank and Richard Strom for their help and
  contributions.}

\newpage

\section*{Appendix}

The following list contains the X and Y coordinates of the 80 antennas
as shown in Fig.~1. Coordinates are in meters, with positive Y to the
north and positive X to the east. The center of the array is at
longitude 21$^{\circ}$23$'$E and latitude 30$^{\circ}$42$'$S.  The
antenna positions are not final and may still change slightly pending
geophysical investigations, but simulations using this layout will
allow investigation of all aspects of the MeerKAT array
\begin{table}[h!]
\tt
\footnotesize
\begin{tabular}{r r | r r}
   X (m)   &    Y (m) &  X (m)  & Y(m)\\
176.061 &  170.880 &      -130.841 &  61.884 \\  
-66.969 &  -606.976 &     82.985 &  -388.329 \\  
242.654 &  132.243 &      -248.388 &  -130.752 \\
164.949 &  -57.655 &      -127.192 &  88.410 \\  
91.389 &  103.574 &       110.910 &  -158.480 \\ 
99.429 &  -334.219 &      -79.387 &  138.975 \\  
-18.404 &  193.350 &      16.065 &  -162.254 \\  
-16.364 &  104.879 &      44.682 &  346.417 \\   
25.054 &  25.450 &        -134.745 &  -221.536 \\
522.297 &  -193.134 &     205.048 &  412.908 \\  
436.562 &  -373.858 &     311.991 &  -72.972 \\  
192.254 &  276.835 &      -5.728 &  -289.321 \\  
-416.962 &  258.853 &     -45.094 &  234.652 \\  
71.097 &  174.981 &       511.999 &  92.634 \\   
198.639 &  -27.787 &      -178.644 &  96.323 \\  
-396.094 &  18.153 &      -449.125 &  -100.602 \\
103.972 &  -39.268 &      1808.727 &  -1849.094 \\
-226.167 &  36.690 &      -1269.759 &  4222.136 \\
-318.506 &  125.535 &     1830.618 &  -872.945 \\
-46.269 &  -2.061 &       1042.711 &  910.623 \\ 
163.962 &  3.157 &        643.621 &  -524.147 \\ 
-389.063 &  -302.250 &    554.371 &  2372.854 \\ 
-229.144 &  345.759 &     -335.516 &  -2204.955 \\
-136.392 &  110.671 &     -311.006 &  -578.926\\
-54.749 &  45.694 &      -85.148 &  -141.128 \\
-213.017 &  282.157 &     4355.955 &  798.831 \\ 
418.688 &  -84.697 &      3220.445 &  2676.747 \\
143.813 &  -63.878 &      2175.319 &  -2864.864 \\
-282.214 &  72.204 &      -3753.578 &  -1390.820 \\
-10.417 &  -244.618 &     388.291 &  -1460.652 \\
194.556 &  83.973 &       1307.771 &  580.398 \\ 
-81.701 &  13.926 &       -3190.450 &  400.925 \\
-429.376 &  -280.397 &    732.985 &  736.581 \\  
234.769 &  -234.833 &     -1510.254 &  122.862 \\
-445.419 &  -177.570 &    -2303.276 &  -209.614 \\
-61.951 &  157.988 &      -766.562 &  474.935 \\
380.132 &  165.606 &      1027.818 &  200.245 \\
94.251 &  66.542 &       -3373.788 &  2233.961 \\
-169.380 &  -136.695 &    -1208.553 &  -3073.213 \\
169.497 &  242.840 &      -980.743 &  -560.740 \\
\end{tabular}
\end{table}

\end{document}